\makeatletter\let\ifGm@compatii\relax\makeatother
\documentclass[pre,aps,twocolumn,showpacs,floatfix,superscriptaddress]{revtex4}
\usepackage{eso-pic,calc}
\usepackage{graphicx}
\usepackage{amsmath, amsthm, amssymb}
\usepackage{epsfig}
\usepackage{latexsym}
\usepackage{bm}

\def\beqr{\begin{eqnarray}}
\def\eqnr{\end{eqnarray}}
\def\beq{\begin{equation}}
\def\bc{\begin{center}}
\def\ec{\end{center}}
\def\eqn{\end{equation}}

\def\rmp#1#2#3{{ Rev. Mod. Phys.} {\bf #1}, #2 (#3)}
\def\prl#1#2#3{{ Phys. Rev. Lett.} {\bf #1}, #2 (#3)}

\def\pre#1#2#3{Phys. Rev. E {\bf #1}, #2 (#3)}

\def\pnas#1#2#3{Proc. Natl. Acad. Sci. (USA) {\bf #1}, #2 (#3)}

\def\physa#1#2#3{Physica A {\bf #1}, #2 (#3)}

\def\sc#1#2#3{Science {\bf #1}, #2 (#3)}

\begin{document}
\title{Exact  solution for a sample space reducing stochastic process}
\author{Avinash Chand Yadav}
\affiliation{School of Physical \& Mathematical Sciences, Central University of Haryana, Mahendergarh 123 029, India}
 
\begin{abstract}
Stochastic processes wherein the size of the state space is changing as a function of time offer models for the emergence of scale-invariant features observed in complex systems. I consider such a sample-space reducing (SSR) stochastic process that results in a random sequence of strictly decreasing integers $\{x(t)\}$,  $0\le t \le \tau$, with boundary conditions $x(0) = N$ and $x(\tau)$ = 1.  This model is shown to be exactly solvable: $\mathcal{P}_N(\tau)$, the probability that the process survives for time $\tau$ is analytically evaluated. In the limit of large $N$, the asymptotic form of this probability distribution is Gaussian, with mean and variance both varying logarithmically with system size: $\langle \tau \rangle \sim \ln N$ and $\sigma_{\tau}^{2} \sim \ln N$. Correspondence can be made between survival time statistics in the SSR process and record statistics of i.i.d. random variables.

\end{abstract}


\maketitle

\section{Introduction}
In many physical systems the state space, the space of all possible outcomes, changes with time. When the size of state space decreases as a function of time,  the process is termed state (or sample) space reducing (SSR).  Examples of such processes are quite common~\cite{Murtra_2015} and instances can be drawn from fields as diverse as material fracture~\cite{Krap_1994, Bohn_2005} to sentence formation or predictive text algorithms in linguistics. The space reduction may be stochastic, and on occasion the size of state space may expand: such systems are referred as noisy SSR processes. Current interest in SSR processes arises from the possibility of such dynamics offering an ``explanation'' of the origin of Zipf's law~\cite{Zipf_1949} or other scale--invariant features observed in complex systems~\cite{Murtra_2015, Newman_2005, Pietronero_2001}. The widespread occurrence of scale invariant features in a variety of natural systems has prompted a number of hypotheses and theories, ranging from multiplicative random process~\cite{Montroll_1982, Amir_2012}, self-organized criticality~\cite{Bak_1987, Bak_1996, Dhar_2006, Yadav_2012}, and, in the context of complex networks, preferential attachment~\cite{Albert_1999}. 

Our interest here is primarily on survival--time statistics in the stochastic SSR process.  Since the SSR problem is exactly solvable, the survival time probability distribution can be easily computed. The mean survival time is an important observable that quantifies physically relevant features of random events that are modelled by a stochastic process with an absorbing boundary condition. Some examples include diffusive search with stochastic resetting~\cite{Majumdar_2011}, diminishing record statistics~\cite{Miller_2013}, and the P{\'o}lya urn process~\cite{Antal_2010}.  

The stochastic SSR process can be described as follows. Denote the size of state space at time $t$ by $x(t)$. 
The time evolution of $x$ can be through a discrete map or via flow equations, depending upon whether variables 
such as the time and the size of state space are discrete or continuous. If both are discrete, for example, then $x$ takes integer values and has the automaton dynamics 
\beq
x(t+1) = G[x(t)],
\eqn  
where $G(x)$ is a random function. At initial time, $x(0) = N$, and $x(t) \ge x(t')$ if $t\le t'$. The process stops at $t= \tau$, when $x(\tau) = 1$. $\tau$ is a random variable that corresponds to the survival time or life span of the process; $x(\tau) = 1$  is a fixed point or an absorbing state  since the state of system cannot change when the size of the state space is 1. The case of continuous time or real-valued processes can be analogously described.

An interesting (and subtle) connection has been noted between the statistics of survival time of the SSR 
process and the {\em record} statistics of independent and identically distributed (iid) random variables~\cite{Schehr_2013}. Recall that an event is termed a record if it betters (or exceeds) all previous instances. The total number of records in a sequence of random variables has been of interest in many applications, and the record statistics of uncorrelated events such as a stochastic time series modelled by iid random variables is well understood. The striking feature in record events of such processes is the existence of power-law distribution for the probability that a record would form after $k$th time step in a sequence of length $N$, namely, $P_N(k) \sim 1/k$~\cite{Wergen_2013}. It is here identified that the survival time of the SSR process is equivalent to the total number of records. The stochastic SSR process is therefore of wider consequence.  

Our main result in this paper is an analytic expression for the survival probability in a model stochastic SSR process. Both the mean as well as the variance of the survival time have a logarithmic dependence on the initial size of the state space. We also present simulation results that are in excellent agreement.

The model that we study is presented in Sec.~II and analytical results for the quantities of interest are also given here. Subsequently numerical results for the survival time statistics of the SSR process are presented in Sec.~III. The equivalence between record statistics of iid random variables and the statistics of the survival times in the SSR process is discussed in Sec.~IV, and the paper concludes with a summary and a brief discussion in Sec.~V.

\begin{figure}[t]
  \centering
  \scalebox{0.4}{\includegraphics{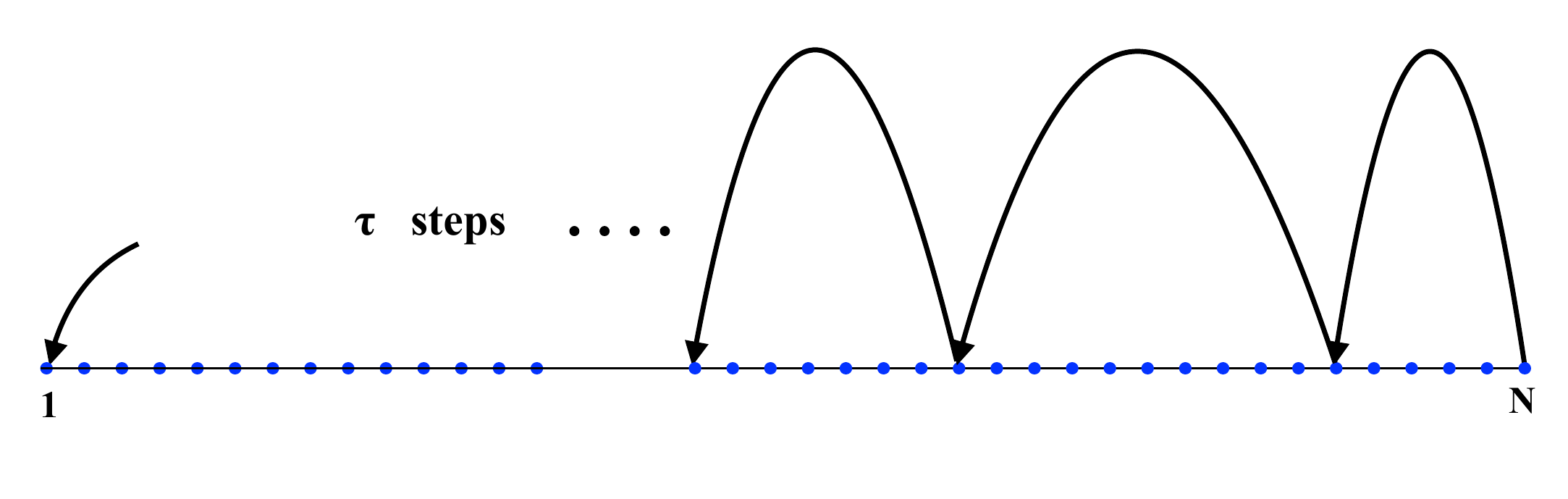}}
  \caption{(Color online) Stochastic SSR process as a directed random hopping with variable step size.}
\label{fig1}
\end{figure}

\begin{figure}[t]
  \centering
  \scalebox{0.5}{\includegraphics{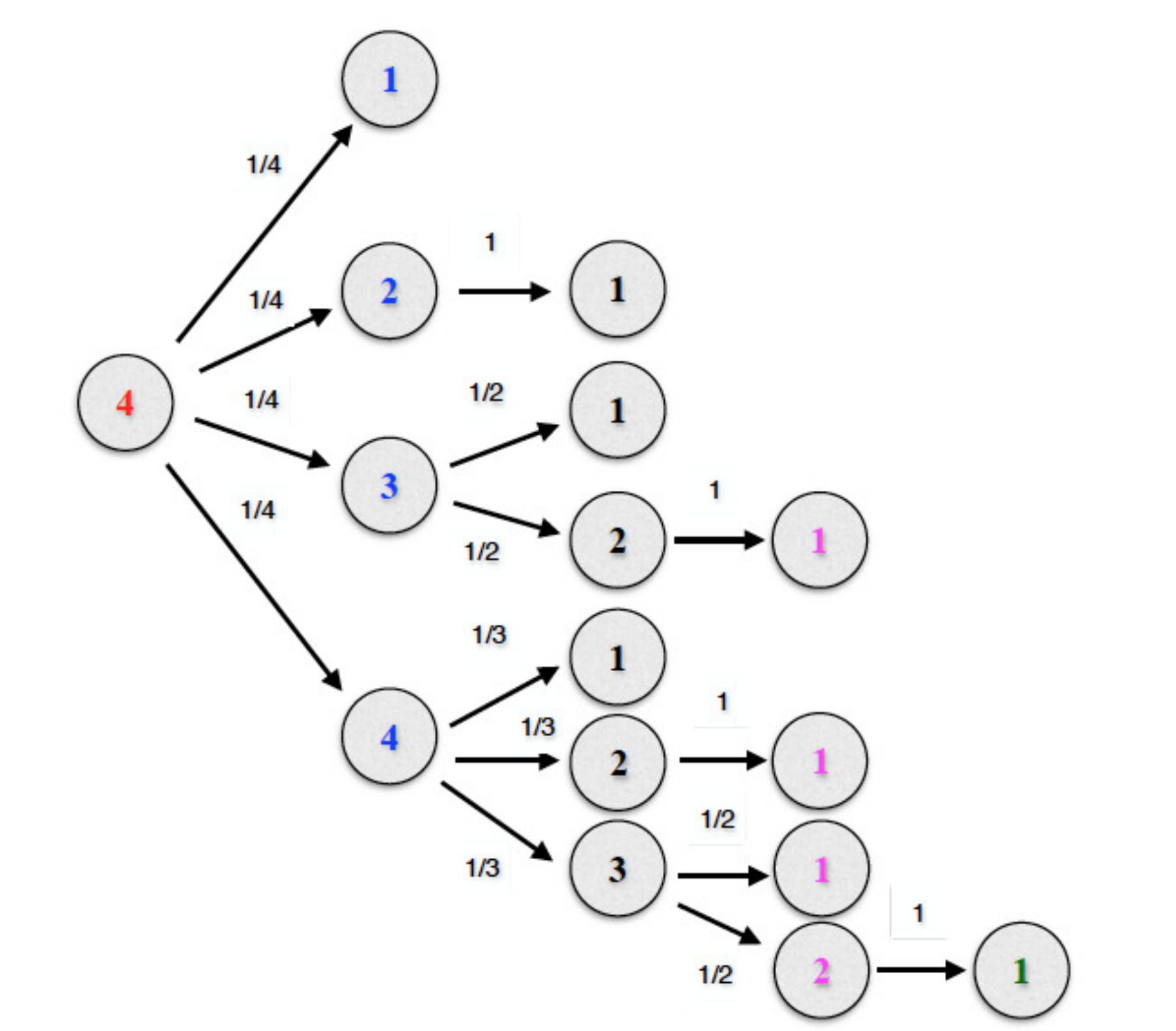}}
  \caption{(Color online) Pictorial representation of all possible paths that originate with system size $N = 4$ for an SSR process. A dice with $N = 4$ faces is thrown at $t = 0$;  the possible outcomes are 1, 2, 3, and 4, each occurring with equal probability 1/4. A typical path might be, for example, $4\to 3\to 2\to 1$ with survival time $\tau = 3$, i.e., the outcome at the first time step is $x(t = 1) = 3$, then a dice with 2 faces will be selected and it is thrown. Now, the possible outcome would be either 1 or 2 with equal probability 1/2. If the outcome is $x(t = 2) = 2$, then finally at next time step the outcome must be $x(t = 3) = 1$ with unit probability.  }
\label{fig2}
\end{figure}

\section{Model Definition and Results} 

The examples discussed in~\cite{Murtra_2015} of stochastic SSRs  can be seen as a directed random hopping process on $\mathbb{Z}^+$, the set of positive integers [see Fig.~\ref{fig1}]. For given  $x(t)$, $x(t+1)$ is a random integer in $[1, x(t)-1]$. The boundary conditions are $x(0)=N$, and $\tau = x^{-1}(1)$, namely the process stops at time
$\tau$ when $x(\tau)$ = 1.  This can be visualised as a sequence of throws of fair dice with $x(t)-1$ faces at  time $t$, with $x(t+1)$ being the outcome of the throw (see Fig.~\ref{fig2}). 

In the set of all possible sequences or paths, the immediate quantities of interest are \begin{enumerate} 
\item The number of paths  $n_\tau$ that  survive for time $\tau$. 
\item The survival probability distribution, namely the probability that a path with survival time $\tau$ occurs, $\mathcal{P}_N(\tau)$, and \item 
The mean survival time $\langle \tau\rangle$, and its variance $\sigma_{\tau}^2$.\end{enumerate}

\subsection{The number of paths that survive up to time $\tau$}
There are total of $2^{N-1}$ possible paths. If $n_{\tau}$ denotes the number of paths with survival time $\tau$, 
\beq
\sum_{\tau} n_{\tau} = 2^{N-1}.
\eqn
It is obvious that $n_1 = 1$ and $n_2 = N-1$, and it can be easily shown recursively that for integer $\tau$, 
\beq n_{\tau} = {{N-1}\choose {\tau-1}}.
\label{ntau}
\eqn 
Dividing by  the total number of paths, namely $2^{N-1}$, the fraction of paths $C_N(\tau)$ that survive upto time $\tau$ is  
\beq
C_N(\tau) = \frac{1}{2^{N-1}}\prod_{i=1}^{\tau-1} \left[\frac{N-i}{i}\right] = \frac{1}{2^{N-1}}\prod_{i=1}^{\tau-1} \left[\frac{1-i/N}{i/N}\right].
\label{eq1}
\eqn
The relation between $C_N(\tau+1)$ and $C_N(\tau)$ is also straightforward to obtain,
\beq
C_N(\tau+1)  = \frac{[1 - \tau/N]}{\tau/N} C_N(\tau),
\label{eq_rec}
\eqn
with boundary conditions $C_N(1) = 1/2^{N-1} = C_N(N)$. Note that the function is symmetric:  $C_N(N-\tau) = C_N(\tau)$.

The Eq.~(\ref{eq1}) can be reexpressed as
\beq
C_N(\tau) = \frac{1}{2^{N-1}}\exp \left\{ \sum_{i=1}^{\tau-1} \left[\ln \left(1- \frac{i}{N}\right) - \ln \frac{i}{N}\right]\right\}.
\eqn
Replacing the summation by an integral, we therefore have
\beq
\frac{1}{N}\ln\left[\frac{C_N(\tau)}{C_N(1)}\right] \approx \int dy[\ln(1-y) - \ln(y)].
\eqn 
Dropping terms independent of $\tau$, one gets 
\beq
\frac{\ln \bar{C}_N(\tau)}{N} \sim -\left[ \left(1-\frac{\tau}{N}\right)\ln \left(1-\frac{\tau}{N}\right) + \frac{\tau}{N}\ln\left(\frac{\tau}{N}\right)\right].\nonumber
\eqn 
and further, in the large $N$ limit, with $\tau/N$ small, keeping the leading order term in the logarithmic expansion, we finally get the relation 
\beq
\frac{\ln \bar{C}_N(\tau)}{N} \sim \left(1-\frac{\tau}{N}\right)\frac{\tau}{N} + ~ {\cal O}(\tau^3). 
\eqn
It is interesting to note that the scaling function has the form of the logistic equation in the variable $u=\tau/N$.

\subsection{The Survival Probability Distribution:  $\mathcal{P}_{N}(\tau)$}

To compute the survival probability distribution we proceed in the following manner. 
Consider the case of $N = 5$ for instance. For $\tau = 1$, $\mathcal{P}_5(1) = \frac{1}{5}$. 
For $\tau = 2$ there would be $n_{\tau} = 4$ terms that need to be summed, giving  $5 \mathcal{P}_5(2) = 
1 + \frac{1}{2} + \frac{1}{3} + \frac{1}{4}$. Working upwards, one can deduce, $\mathcal{P}_5(5) = 
{1\over{5!}}$. In general, therefore, one notes that $\mathcal{P}_N(1) = \frac{1}{N}$; $\mathcal{P}_N(N) = 
\frac{1}{N!}$, and
\beq
N\mathcal{P}_N(\tau) = \sum_{i=1}^{n_{\tau}}\mathbb{T}_i(\tau) = \sum_{i=1}^{n_{\tau}}\prod_{k=1}^{\tau-1}p(k).
\label{eq2}
\eqn
{with $p \in \{1/1, 1/2, 1/3, \dots, 1/(N-1)\}$. Here the $i$th term $\mathbb{T}_i(\tau)$ represents contribution from one of the path that survives for time $\tau$ out of $n_{\tau}$ paths, and this is equal to one of the all different possible combinations of product of $\tau-1$ distinct factors of $p$. As $\tau-1$ entries, out of $N-1$ possible values of $p$, have to be involved in writing the $i$th term $\mathbb{T}_i(\tau)$, we immediately note that the total number of different combinations are  $n_{\tau}$ [see Eq.~(\ref{ntau})]}

For large $N$, the mean value of $p$ is 
\beq
\langle p \rangle = \frac{1}{N}\sum_{i=1}^{N}\frac{1}{i} = \frac{1}{N}H_N,
\eqn
where $H_N$ is the Harmonic number $H_N = \ln N + \gamma + \mathcal{O}(N^{-1})$ with $\gamma = 0.57721\dots$ 
being Euler's constant. Then, the average survival time would be 
\beq
\langle \tau \rangle \sim N\langle p \rangle \sim \ln N.
\eqn

Alternatively, we can also write the survival probability as a sum of probabilities, with time $\tau - 1$, in terms of lower values of $N$ as
\beq
N\mathcal{P}_N(\tau) = \sum_{j=\tau-1}^{N-1}\mathcal{P}_j(\tau-1),
\eqn
and this can be further expressed as a recursion relation 
\beq
N\mathcal{P}_{N}(\tau) = (N-1)\mathcal{P}_{N-1}(\tau) + \mathcal{P}_{N-1}(\tau-1),
\label{spr}
\eqn
for $1<\tau \le N$ and $N>1$ with $N\mathcal{P}_{N}(1) = 1$. From numerical implementation point of view the  recursion relation in Eq.~(\ref{spr}) is the most useful.

Further, in the continuum limit of $N$, the difference equation~(\ref{spr}) can be expressed as a differential equation 
\beq
\frac{d\mathcal{P}_{N}(\tau)}{dN} = \frac{1}{N}\left[ \mathcal{P}_{N}(\tau-1) - \mathcal{P}_{N}(\tau)\right].
\eqn  
In order to solve this, we introduce a generating function 
\beq
f(N,z) = \sum_{\tau=1}^{N}\mathcal{P}_{N}(\tau) z^{\tau-1},
\label{gf}
\eqn
and then in terms of the generating function the differential equation reduces as
\beq
\frac{\partial f(N,z)}{\partial N} = \frac{1}{N}(z-1)f(N,z),
\label{gfdeq}
\eqn
with initial condition $f(1,z) = 1$, giving solution 
\beq
f(N,z) = \exp{[(z-1)\ln N]}.
\label{gf1}
\eqn

With the help of this generating function, it is easy to see that the normalization is $f(N,1) = \sum_{\tau}P_{N}(\tau) = 1$, the mean survival time is 
\beq
\langle \tau-1 \rangle = \frac{\partial f(N,z)}{\partial z}\bigg|_{z=1} = \ln N,
\eqn 
and the second factorial moment is 
\beq
\langle (\tau-1)(\tau-2) \rangle =  \frac{\partial^2 f(N,z)}{\partial z^2}\bigg|_{z=1}  = (\ln N)^2.
\eqn
Thus, the variance is given as 
\beq
\sigma_{\tau}^{2}(N) = \langle \tau^2 \rangle - \langle \tau \rangle^2 \sim \ln N.
\eqn
Using Eqs.~(\ref{gf}) \& (\ref{gf1}), an expression for the survival probability can be written as
\beq
\mathcal{P}_{N}(\tau) = \frac{1}{(\tau-1)!}\frac{\partial^{\tau-1}f(N,z)}{\partial z^{\tau-1}}\bigg|_{z = 0} =
\frac{(\ln N)^{\tau-1}}{(\tau-1)!N}.
\label{sd_pd}
\eqn
The survival time is a discrete random variable with Poisson probability distribution, but note that $\tau \in \{1, 2 \dots, N\}$ and the equation~(\ref{gfdeq}) is only consistent when the upper limit of the survival time, i.e., $N$ is large.  Also, note that the Poissonian form of the survival time distribution with parameter $\ln N$ [see Eq.~(\ref{sd_pd})] reduces to normal distribution for a shifted and scaled variable $v = (\tau - \langle \tau \rangle )/\sigma_{\tau}$, when $N$ is large.

\begin{figure}[b]
  \centering
  \scalebox{0.65}{\includegraphics{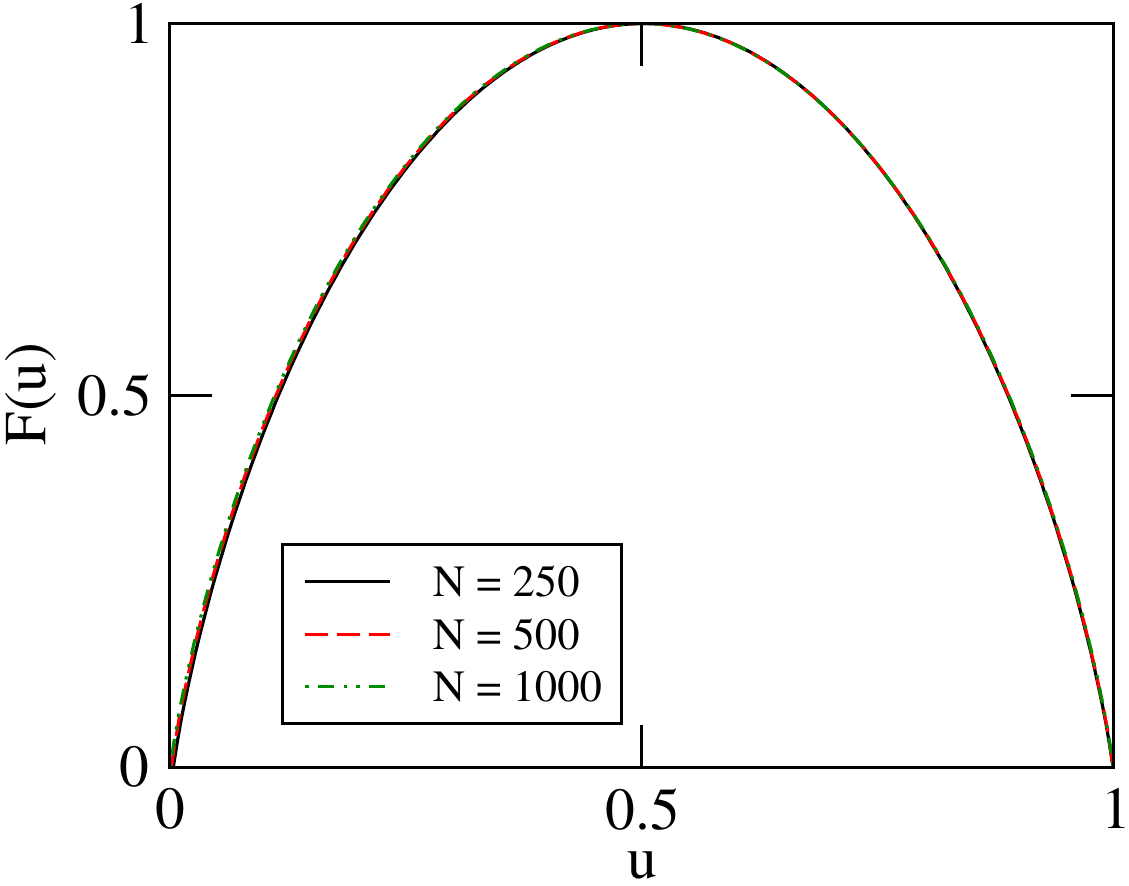}}
  \caption{(Color online) Data collapse curves for $n_{\tau}$: plot of $F(u) = \ln{\bar{C}_N(\tau)}/N$ with $u = \tau/N$.  }
\label{fig3}
\end{figure}

\section{Numerical Results} 
Equation (\ref{eq_rec}) is used to compute $\bar{C}_N(\tau)$ and hence $F(u) = \ln{\bar{C}_N(\tau)}/N$ as a function of $\tau$ for different values of $N = 250, 500,$ and 1000. In Fig.~\ref{fig3} we show a plot of data collapse between $F(u)/F_m$, where $F_m$ is the maximum value of $F(u)$, and the scaled time $u = \tau/N$. Then, we also numerically compare this data collapse curve with the expected theoretical form of scaling function $F(u)\sim u(1-u)$, upto the first order. We observe that these two curves don't show a good agreement, suggesting that higher order corrections to the scaling function dominate.

\begin{figure}[t]
  \centering
  \scalebox{0.65}{\includegraphics{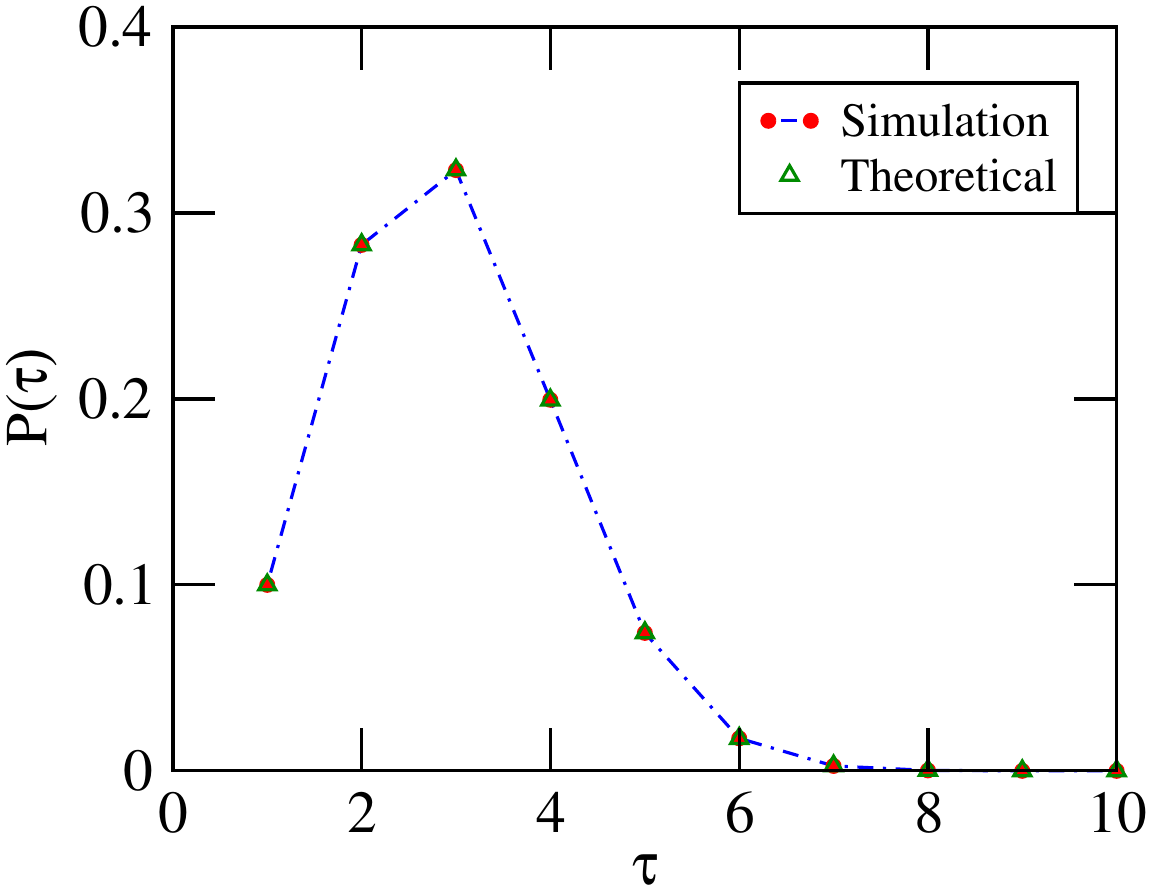}}
  \caption{(Color online) Survival probability distribution:  Shaded circle denotes data obtained from simulation with $N = 10$ and total number of samples $10^8$, and open triangle corresponds to theoretical curve computed using Eq.~(\ref{spr}).}
\label{fig4}
\end{figure}

\begin{figure}[t]
  \centering
  \scalebox{0.65}{\includegraphics{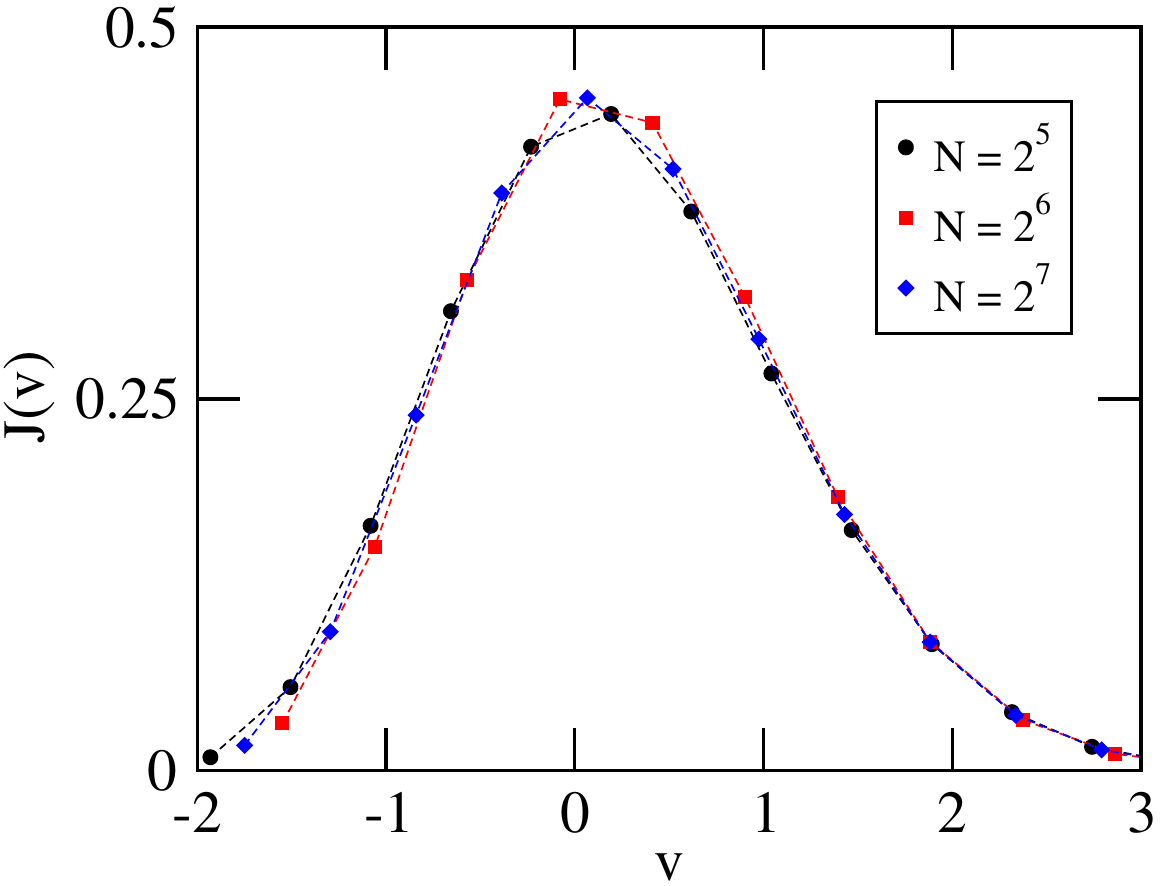}}
  \caption{(Color online) Data collapse curves for the survival probability distribution with different system size $N$ calculated using Eq.~(\ref{spr}). Here $v = (\tau-\ln N)/\sqrt{\ln N}$ and $J(v) = \sqrt{\ln N}P_{N}(\tau)$.}
\label{fig5}
\end{figure}

Monte Carlo simulations have been performed to realize the SSR process, independently $10^8$ times. Then we compute the normalized survival probability distribution $\mathcal{P}_N(\tau)$. In Fig.~\ref{fig4} we show a plot of the survival probability distribution computed using theoretical expression [see Eq.~(\ref{spr})] and simulations both.   We find a good agreement between simulation and theoretical results. The behaviour of $\mathcal{P}(\tau)$ with different system size $N$ computed using Eq.~(\ref{spr}) is shown in Fig.~\ref{fig5}.

\section{A map between the survival time statistics of the SSR process and the record statistics of iid random variables}
Consider $N$ iid random variables $\{X_i\}$ drawn from a continuous probability distribution. $k$th entry forms an upper or lower record if this is the largest or smallest number with respect to existing $k-1$ entries. Let us use an indicator variable $\xi_k$ that denotes 1 if the $k$th entry forms a record, zero otherwise.  Then the total number of records  out of $N$ entries would be $R_N = \sum_{k=1}^{N}\xi_k$. The mean number of records is calculated as $\langle R_N\rangle = \sum_{k=1}^{N}\langle \xi_k\rangle$, where $\langle \xi_k\rangle$ denotes the rate at which a record would occur.  Since each entries are independent and equally probable, the $k$th entry is a record with rate $\langle \xi_k\rangle = 1/k$. For large $N$, the probability distribution of $R_N$ behaves as Gaussian distribution with mean and variance both varying as $\ln N$~\cite{Schehr_2013}.

\begin{figure}[t]
  \centering
  \scalebox{0.65}{\includegraphics{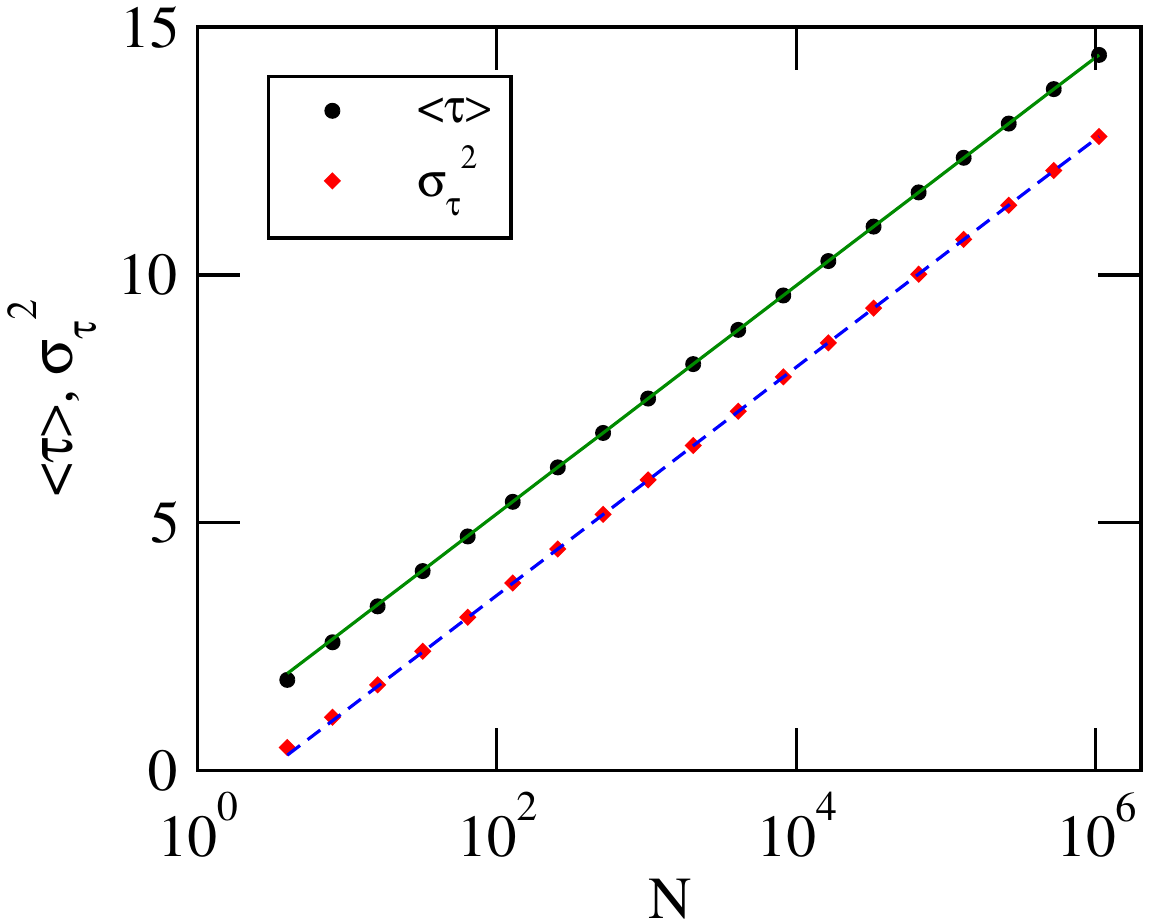}}
  \caption{(Color online) The plots show the mean and variance of survival time that is $\langle \tau \rangle$ and $\sigma_{\tau}^{2}$ as a function of system size $N$. Each point is obtained averaging over $10^8$ independent realizations of the SSR process. $N$ is chosen in steps $2^i$ where $i$ runs from 2 to 20. The difference of mean and variance is $\langle \tau \rangle - \sigma_{\tau}^{2} \approx 1.647$ for large $N = 2^{20}$, and this number is approximately equal to $\zeta(2) = \pi^2/6 \approx 1.646$, where $\zeta(n)$ is the  Riemann Zeta function. The straight lines, corresponding to the mean and variance, are computed using theoretical form [see Eq.~(\ref{msd})].}
\label{fig6}
\end{figure}

In order to see a connection between the records statistics of iid random variables and the SSR process, we can consider the survival time as the total number of records, i.e. $\tau \equiv R_N$. In bouncing ball representation if the ball jumps from site $x(j)$ to a site $x(k)$, the  indicator $\xi_k$ is set on, i.e., 1 and off, i.e., 0 for sites where the ball did not visit. The survival time can be expressed as a sum of all values of indicator variable. The rate or transition probability or the visiting probability that at $k$th step the ball jumps is $P_N(k) = \langle \xi_{k}\rangle = 1/k$~\cite{Murtra_2015}.  Clearly, the mean survival time $\langle \tau \rangle = \sum_{k=1}^{N}\langle \xi_k\rangle$ grows as $\ln N$. The connection  with the records statistics of iid random variables indicates that for large $N$, the asymptotic form of the  survival time probability distribution behaves as Gaussian with mean and variance both varying as $\ln N$. More precisely, we have 
\beqr
\langle \tau \rangle = \ln N + \gamma + \mathcal{O}(N^{-1}),\nonumber\\ 
\sigma_{\tau}^{2} = \ln N + \gamma -\zeta(2) + \mathcal{O}(N^{-1}),
\label{msd}
\eqnr
where $\zeta(2) = \pi^2/6$, and 
\beq
\mathcal{P}_{N}(\tau) = \frac{1}{\sqrt{2\pi \ln N}}\exp\left[ -\frac{(\tau - \ln N)^2}{2 \ln N}\right]. 
\eqn
Note that this expression is valid for large $N$ and treating $\tau$ as a continuous variable. Results shown in Fig.~\ref{fig6} provide a clear evidence that the statistics of survival time of the SSR process and the record statistics of iid random variables behave exactly in the same manner.

\section{Summary and Discussion}

We have studied  the statistics of survival time, an important physical observable for the SSR process. This is an interesting model to understand systems where the size of state space changes with time, and provides an explanation for the Zipf's law. The analytical tractability of the model allows us to exactly calculate the probability distribution of survival time. We have also checked these features through simulations that agree well with the theoretical results. A map between the survival time statistics of the SSR process and the record statistics of iid random variables has been found. The mean and variance of the survival time grows logarithmically as a function of system size.

It would be further useful to explore a comparison with cases where mean path grows logarithmically with system size. It is interesting to note that the average number of divisors of integers in the interval $[1, N]$ asymptotically behaves as $\sim \ln N + 2\gamma -1$. In order to compute the total number of divisors of an integer $N$, including 1 and the number itself, an indicator variable is used 
\beq
\xi_k = \left\lfloor \frac{N}{k} \right\rfloor- \left \lfloor\frac{N-1}{k}\right\rfloor,
\eqn
where $\lfloor \cdot \rfloor$ is the integer part of its argument. Then the total number of divisors is $d(N) = \sum_{k=1}^{N}\xi_k$, and the average number of divisors would be $\langle d(N) \rangle = \sum_{k=1}^{N} \langle \xi_k\rangle$, where the rate behaves as $\langle \xi_k\rangle \sim 1/k$~\cite{Luque_2008, Schroeder_1997}. This suggests that there may be a relation with the SSR process.

The mean path in complex networks with $N$ nodes grows as $\langle l \rangle \sim \ln N/\ln\langle k \rangle$, where $\langle k \rangle$ is the average degree of each node~\cite{Albert_2002, Redner_2005}. Also, the statistics of cycles in a random permutation with uniform measure is equivalent to the record statistics of iid random variables~\cite{Flajolet_2009}. We note that the survival time for the SSR process is equivalent to the number of cycles in a random permutation of $N$ objects, or the total number of records out of $N$ samples of iid random variable, or the path length of the complex network. The wide connections of the SSR process and its analytical tractability suggest that further investigation along these lines would be fruitful.

\section*{ACKNOWLEDGMENTS}

ACY thanks R. Ramaswamy for valuable discussions on this topic and critical reading of the manuscript.

\end{document}